\documentclass[aps,prl,twocolumn,superscriptaddress]{revtex4-1}
\usepackage{amssymb, amsmath}
\usepackage{graphicx,onlyamsmath}
\usepackage[normalem]{ulem}  

\usepackage{natbib}
\usepackage{xcolor}

\begin{document}

\title{Temperature-induced topological phase transition in HgTe quantum wells}

\author{A.~M.~Kadykov}
\affiliation{Laboratoire Charles Coulomb, UMR Centre National de la Recherche Scientifique 5221, University of Montpellier, F-34095 Montpellier, France.}
\affiliation{Institute for Physics of Microstructures RAS, GSP-105, Nizhni Novgorod 603950, Russia}

\author{S.~S.~Krishtopenko}
\thanks{These two authors contributed equally.}
\affiliation{Laboratoire Charles Coulomb, UMR Centre National de la Recherche Scientifique 5221, University of Montpellier, F-34095 Montpellier, France.}
\affiliation{Institute for Physics of Microstructures RAS, GSP-105, Nizhni Novgorod 603950, Russia}

\author{B.~Jouault}
\thanks{These two authors contributed equally.}
\affiliation{Laboratoire Charles Coulomb, UMR Centre National de la Recherche Scientifique 5221, University of Montpellier, F-34095 Montpellier, France.}

\author{W. Desrat}
\affiliation{Laboratoire Charles Coulomb, UMR Centre National de la Recherche Scientifique 5221, University of Montpellier, F-34095 Montpellier, France.}

\author{W.~Knap}
\affiliation{Laboratoire Charles Coulomb, UMR Centre National de la Recherche Scientifique 5221, University of Montpellier, F-34095 Montpellier, France.}

\author{S. Ruffenach}
\affiliation{Laboratoire Charles Coulomb, UMR Centre National de la Recherche Scientifique 5221, University of Montpellier, F-34095 Montpellier, France.}

\author{C. Consejo}
\affiliation{Laboratoire Charles Coulomb, UMR Centre National de la Recherche Scientifique 5221, University of Montpellier, F-34095 Montpellier, France.}

\author{J.~Torres}
\affiliation{Institut d'Electronique et des Systemes, UMR Centre National de la Recherche Scientifique 5214, University of Montpellier, F-34095 Montpellier, France.}

\author{S.~V.~Morozov}
\affiliation{Institute for Physics of Microstructures RAS, GSP-105, Nizhni Novgorod 603950, Russia}

\author{N.~N.~Mikhailov}
\affiliation{Institute of Semiconductor Physics, Siberian Branch, Russian Academy of Sciences, pr. Akademika Lavrent'eva 13, Novosibirsk, 630090 Russia}
\affiliation{Novosibirsk State University, Pirogova st. 2, 630090 Novosibirsk, Russia.}

\author{S.~A.~Dvoretskii}
\affiliation{Institute of Semiconductor Physics, Siberian Branch, Russian Academy of Sciences, pr. Akademika Lavrent'eva 13, Novosibirsk, 630090 Russia}
\affiliation{Novosibirsk State University, Pirogova st. 2, 630090 Novosibirsk, Russia.}

\author{F.~Teppe}
\email[]{frederic.teppe@umontpellier.fr}
\affiliation{Laboratoire Charles Coulomb, UMR Centre National de la Recherche Scientifique 5221, University of Montpellier, F-34095 Montpellier, France.}
\date{\today}

\begin{abstract}
We report a direct observation of temperature-induced topological phase transition between trivial and topological insulator in HgTe quantum well. By using a gated Hall bar device, we measure and represent Landau levels in fan charts at different temperatures and we follow the temperature evolution of a peculiar pair of "zero-mode" Landau levels, which split from the edge of electron-like and hole-like subbands. Their crossing at critical magnetic field $B_c$ is a characteristic of inverted band structure in the quantum well. By measuring the temperature dependence of $B_c$, we directly extract the critical temperature $T_c$, at which the bulk band-gap vanishes and the topological phase transition occurs. Above this critical temperature, the opening of a trivial gap is clearly observed.
\end{abstract}

\maketitle

The first two-dimensional (2D) system, in which a topological insulator (TI) phase was predicted~\cite{Q1} and then experimentally observed~\cite{Q2}, was HgTe/Cd$_{x}$Hg$_{1-x}$Te quantum wells (QWs) with inverted band structure. The inversion of electron-like level \emph{E}1 and hole-like level \emph{H}1 induces spin-polarized helical edge states~\cite{Q2,Q16,Q17}. When \emph{E}1 and \emph{H}1 levels cross, the band structure mimics a linear dispersion of massless Dirac fermions~\cite{Q3} corresponding to the phase transition between normal band insulator (NI) and TI states.

The QW thickness $d$ was earlier employed as a tuning parameter of these different quantum phases of matter. The crossing of the energy bands arising when $d$ equals a critical value $d_c$ ($d_c\approx6.3$ nm for $x$=0.7 and QWs grown on CdTe buffer~\cite{Q3}) was measured and identified as a key signature for the topological phase transition~\cite{Q2,Q3}. However, in addition to the QW thickness, temperature~\cite{Q4,Q5} and hydrostatic pressure~\cite{Q6} also induces the transition between NI and TI phases across the critical gapless state. The temperature effect on the band ordering in HgTe/CdHgTe QWs is mainly caused by a strong temperature dependence of the energy gap at the $\Gamma$ point of the Brillouin zone between the $\Gamma_6$ and $\Gamma_8$ bands in HgCdTe crystals~\cite{Q7}.


Under magnetic field, a significant way of discriminating the TI phase associated with the helical edge states of the Quantum Spin Hall Effect (QSHE), and the NI phase presenting the chiral edge states of the ordinary Quantum Hall Effect (QHE), is to probe the behavior a particular pair of Landau levels (LLs), called zero-mode LLs~\cite{Q2}. These zero-mode LLs split with magnetic field from the \emph{E}1 and \emph{H}1 subbands and correspond to the lowest LL of the conduction band and the highest LL of the valence band. In the TI phase, these \emph{E}1 and \emph{H}1 subbands are inverted and intersect at the sample boundaries to create the helical edge channels of the QSHE. In this case, the corresponding zero-mode LLs cross at a critical magnetic field $B_c$, above which the inverted band ordering is transformed into the normal one~\cite{Q2}.

In the NI phase, in which \emph{E}1 subband lies above \emph{H}1 subband, the zero-mode LLs never cross. This can be conditionally interpreted as negative values for $B_c$. Thus, $B_c=0$ corresponds to a topological phase transition between NI and TI phases. As it is for the band ordering, a critical magnetic field also depends on temperature and pressure, and therefore, can be varied by tuning these external parameters~\cite{Q6}.


Recently, Wiedmann {\it et al.}~\cite{Q5}, by analyzing magnetotransport data at high and low temperatures, have experimentally shown that the TI phase is destroyed at high temperature. However, due to high values of the critical temperature $T_c$ (more than 200~K) corresponding to the topological phase transition, the presence of critical gapless state was not directly observed. Another work~\cite{Q13} has reported a temperature evolution of the band-gap observed by far-infrared LL spectroscopy. Although the critical temperature was found to be 90~K, an inadequate position of the Fermi level, combined with the inability to vary the carrier density in the samples, hindered a clear observation of the topological phase transition.

In this work, we report on the first clear observation of topological phase transition induced by temperature. Accurate values of $B_c$ and a temperature phase diagram have been extracted from LL fan charts based on magnetotransport measurements, as firstly performed by B\"{u}ttner {\it et al.}~\cite{Q3} for different QW widths at 4.2~K. By following temperature dependence of $B_c$, we define a critical temperature $T_c$, at which the gapless state arises. To describe our experimental results, realistic band structure calculations, based on the eight-band Kane Hamiltonian with temperature-dependent parameters~\cite{Q6}, have been performed.

The QW studied in this work was grown by molecular beam epitaxy (MBE) on a [013]-oriented semi-insulating GaAs substrate with a relaxed CdTe buffer~\cite{Q11}. The HgTe QW of 6.5~nm width was embedded in 40-nm Cd$_{0.65}$Hg$_{0.35}$Te barriers (critical thickness $d_c=6.2$~nm, see also Fig.~\ref{Fig:4}). A 40-nm CdTe cap layer was deposited on top of the structures. The barriers from both sides of the QW were selectively doped with indium resulting in a 2D electron concentration of a few $10^{11}$ cm$^{-2}$ at low temperatures. After MBE growth, 100 nm SiO$_2$ and 200 nm Si$_3$N$_4$ dielectric layers were deposited on top of the structure by a plasmochemical method. The gated Hall bar has a total length of 650~$\mu$m and a total width of 50~$\mu$m. Magnetotransport measurements have been performed in a variable temperature insert with a base temperature $T=1.7$~K equipped with a superconducting coil. All the measurements have been done with a current of 1~nA, to avoid heating.

\begin{figure}
\includegraphics [width=1.0\columnwidth, keepaspectratio] {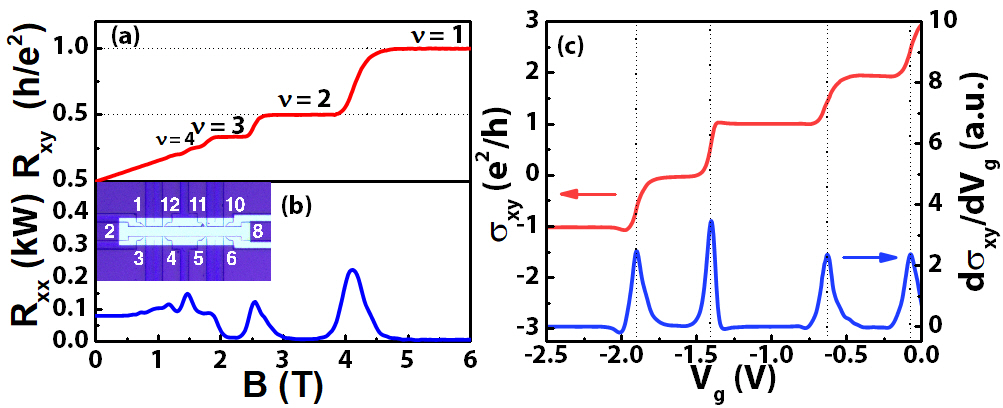} 
\caption{\label{Fig:1} Hall resistance $R_{xy}$ (a) and Shubnikov-de Haas (SdH) oscillations, $R_{xx}=R_{28,34}$, (b) at 1.7 K at zero gate voltage. The insert schematically shows a picture of the Hall bar. (c) Hall conductivity $\sigma_{xy}$ and its derivative $\partial\sigma_{xy}/\partial V_G$ as a function of gate voltage at 1.7 K and magnetic field of 2.4 T.}
\end{figure}

Figure~\ref{Fig:1}(a,b) presents Hall resistance and Shubnikov-de Haas (SdH) oscillations at 1.7~K. The Hall resistance shows pronounced plateaus at both even and odd multiples of $h/e^2$. The Hall conductivity as a function of the gate voltage $V_G$ at 1.7~K and magnetic field of 2.4~T, as well as its derivative $\partial\sigma_{xy}/\partial V_G$, are both shown in the panel (c). Each peak corresponds to the crossings of the Fermi level with the given LL. This method allows for a visualization of LLs from \emph{E}1 and \emph{H}1 subbands and clear reconstruction of their behaviour in magnetic field.

\begin{figure}
\includegraphics [width=1.0\columnwidth, keepaspectratio] {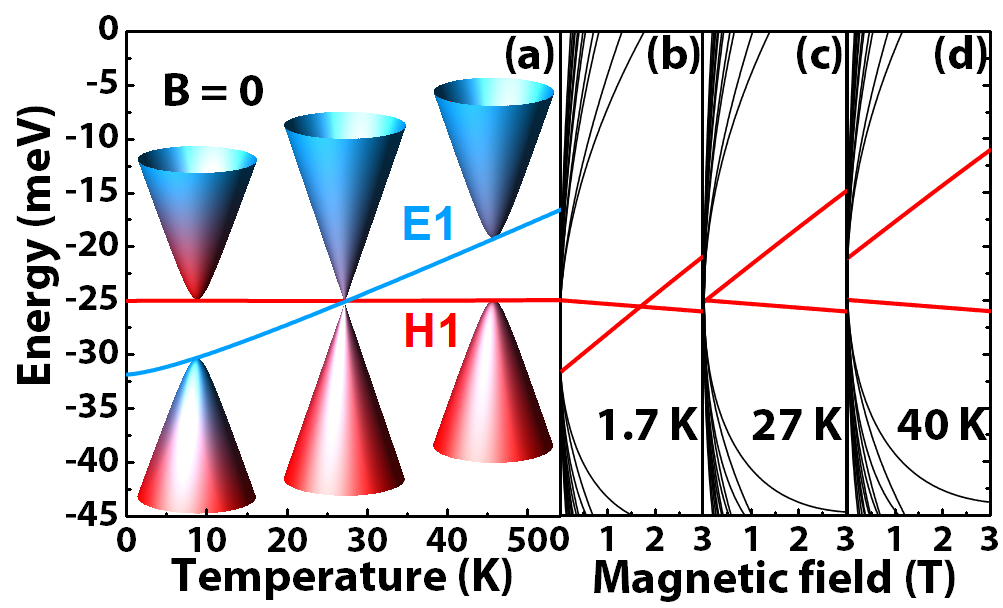} 
\caption{\label{Fig:2} (a) Evolution of \emph{E}1 and \emph{H}1 subbands (at $\mathbf{k}=0$) and band structure in the sample with temperature. Blue and red colored shading indicates contribution from electron-like and hole-like states at given quasimomentum $\mathbf{k}$ point. (b-d) Landau level fan chart at different temperatures: 1.7 K (TI phase), 27 K (critical gapless state) and 40 K (NI phase). A pair of zero-mode LLs is presented by red curves.}
\end{figure}

Figure~\ref{Fig:2}(a) provides a temperature evolution of \emph{E}1 and \emph{H}1 subbands at zero quasimomentum and Fig.~\ref{Fig:2}(b-d) show the corresponding LL fan charts at three different temperatures, evidencing the position of the zero mode LLs. The upper zero-mode LL level split from the conduction band has a pure heavy-hole character and its energy decreases with $B$. By contrast,  the second zero-mode LL starting from valence band has an electron component~\cite{Q2} and its energy increases with $B$.

At $T \simeq 1.7$ K, the $E$1 subband is well below the $H$1 subband and the two corresponding zero-mode LLs cross at a positive magnetic field value $B_c \simeq 1.6$ T. This positive $B_c$ characterizes the TI phase. By increasing the temperature, the width of the TI gap decreases, yielding that  $B_c$ decreases. At a critical temperature $T_c = 27$ K, the zero-mode LLs intersect at $B_c \simeq 0$ T, which means that the subbands $E1$ and $H1$ have the same energy and therefore the topological gap closes. At $T_c$, the system hosts single-valley massless Dirac fermions~\cite{Q3}. Above $T_c$, the crossing of the zero-mode LLs can be extrapolated to a negative value of magnetic field.


To demonstrate experimentally a temperature driven phase transition in our samples, we reconstruct the LL fan chart by means of magnetotransport, measured in a wide range of gate voltages at temperatures from 1.7~K up to 40~K. As mentioned above, the derivative of the Hall conductivity has it maximum values when the Fermi level crosses one of the LLs. Therefore, plotting $\partial\sigma_{xy}/\partial V_G$ for each magnetic field values makes possible to visualize LL fan chart~\cite{Q3}. This allows for accurate extraction of the critical magnetic field $B_c$,
at which the zero-mode LLs cross. Since $B_c$ is related to the changing of band ordering, the measurement of LL fan charts performed by tuning the temperature is an efficient tool to probe a temperature-induced phase transition between NI and TI phases.

\begin{figure}
\includegraphics [width=1.0\columnwidth, keepaspectratio] {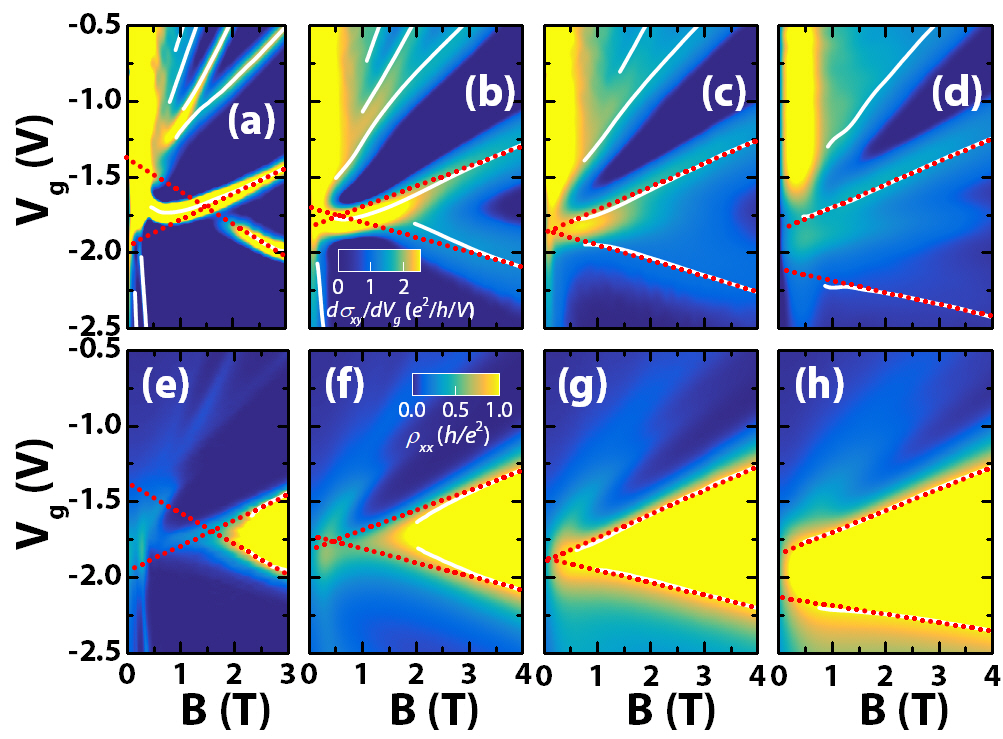} 
\caption{\label{Fig:3} (a-d) Colormap of $\partial\sigma_{xy}/\partial V_g$ as a function of both magnetic field and gate voltage, at 1.7~K, 20~K, 27~K and 40~K respectively. The white solid curves correspond to $\sigma_{xy}= (n+1/2) e^2/h$. In the NI state ($B>2.2$ T), the isolines of $\sigma_{xy}= \pm e^2/2h$ are linearly fitted by the red dotted lines giving the values of $B_c$. (e-h) The false color maps show the longitudinal resistivity at the same temperatures as in the top panels. The white curves correspond to the isolines of $\rho_{xx}=h/e^2$. The red dotted lines are the linear fits of these curves in the field range of $B=$ 2.2--6~T. The crossing point of the two red lines gives another estimate for $B_c$.}
\end{figure}


The colormaps in Fig.~\ref{Fig:3}(a-d) were obtained by plotting the derivative $\partial\sigma_{xy}/\partial V_g$ as a function of $B$, at different temperatures. The two zero modes LLs are observed separately above $B \simeq 1.5$ T. They seem to emerge from the resistance maximum at $V_g= -1.7$~V, $B=0$ T. The $\partial\sigma_{xy}/\partial V_g$ trace of the zero-mode LL, originating from the \emph{H}1 subband, obviously broadens and fades out at large $T$, probably due to strong temperature dependence of mobility caused by large effective mass of holes. It becomes hardly distinguishable above 20~K. However, the zero mode LLs also coincide with $\sigma_{xy}= \pm e^2/2h$ and $\rho_{xx} \simeq h/e^2$ (from the semicircle relation $\sigma_{xx}^2+(\sigma_{xy} \mp 1/2)^2= 1/4$)~\cite{Q32}).

In Fig.~\ref{Fig:3}(a-d), the white curves correspond to $\sigma_{xy}=(n+1/2)e^2/h$, where $n$ is an integer. They are clearly defined up to 40~K and underline the LL positions, in agreement with the $\partial\sigma_{xy}/\partial V_g$ maxima. The estimated position of the two zero mode LLs do not follow the lines of constant filling factor, which implies that LLs overlap. Numerical simulations~\cite{SM}, performed by taking into account a LL broadening $\Gamma \simeq 4$ meV evidence that $B_c$ can be estimated from the linear extrapolation of the LL position in the NI gap, toward lower magnetic fields. The linear interpolations  of the two curves $\sigma_{xy}=\pm  e^2/2h$ for $B> 2.2$ T are marked by the red dotted lines. At $T=1.7$~K, these lines cross at a finite magnetic field and give an estimate $B_c\approx1.5$ T. As $T$ increases, the crossing goes to lower magnetic fields and it finally vanishes at $T\simeq 27$~K. The crossing of the zero-mode LLs at zero magnetic field gives a direct indication of a critical gapless state~\cite{Q3}, revealing a topological phase transition at $T\simeq 27$~K. At higher temperatures ($T=40$~K, see Fig.~\ref{Fig:3}(d)), the crossing of the zero-mode LLs can be extrapolated to a negative value of magnetic field, evidencing a NI phase with trivial band ordering.

To check the last point, we also plot the longitudinal resistivity in color-coded graphs as a function of $B$ and $V_g$ at temperatures up to 40~K, see the bottom panels in Fig.~\ref{Fig:3}. The whole set of LLs is not seen here as in the top panels, but the traces of both zero-mode LLs are clearly visible at the edges of the main $\rho_{xx}$ peak. The red dotted lines correspond to the linear fitting of the isolines of $\rho_{xx}=h/e^2$ in the field range of $B=$ 2.2--6~T. The crossing of the fitting lines in lower fields gives another estimate of $B_c$.

\begin{figure}
\includegraphics [width=1.0\columnwidth, keepaspectratio] {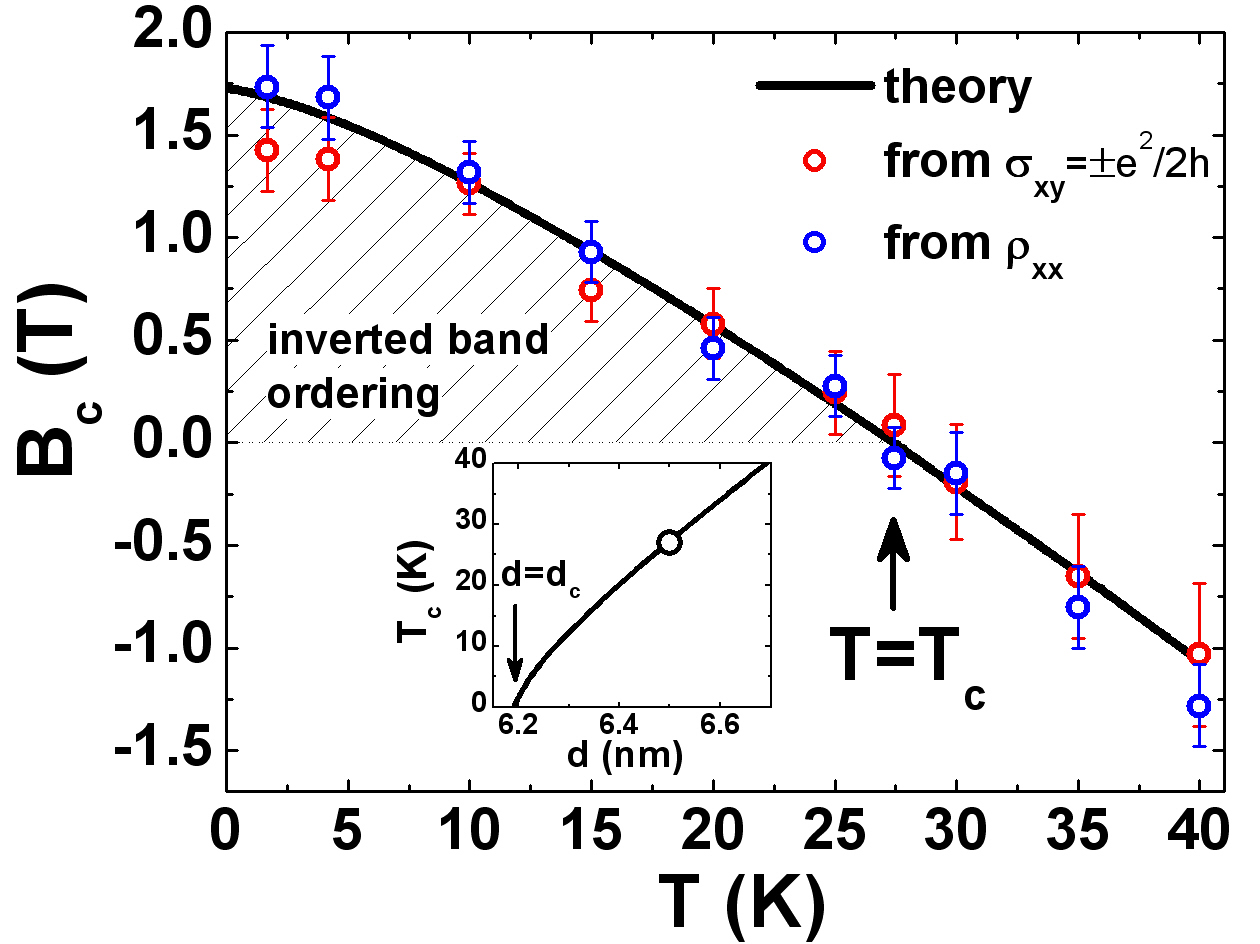} 
\caption{\label{Fig:4} Theoretical (solid curve) and experimental (open symbols) values of critical magnetic field $B_c$ as a function of temperature. The values marked by the red symbols are obtained from the data set in the top panels of Fig.~\ref{Fig:3}. The blue symbols are marked the values of $B_c$ extracted from $\rho_{xx}$. The sparse and white regions correspond to the inverted and trivial band ordering, respectively. The inset shows $T_c$ as a function of $d$ in (013) HgTe/Cd$_{0.65}$Hg$_{0.35}$Te QWs grown on CdTe buffer. The open symbol represents experimental value measured in our sample.}
\end{figure}

Figure~\ref{Fig:4} provides a comparison of experimental and theoretical values of critical magnetic field.  Details of theoretical calculations can be found elsewhere~\cite{Q6}.
Additional analysis based on a simplified Dirac-like Hamiltonian shows that temperature evolution of $B_c$ is caused by temperature dependence of the band gap in our sample (see Fig.~\ref{Fig:2}), while temperature effect on the dispersion of the zero-mode LLs is negligibly small~\cite{SM}. Two sets of experimental values have been obtained from analysis of experimental data presented in the top and bottom panels of Fig.~\ref{Fig:3}. For each set, the experimental precision for $B_c$ is the (1$\sigma$) standard deviation given by the $\rho_{xx}$ and $\sigma_{xy}$ estimates. It is seen that experimental values of $B_c$ from different sets reproduce quantitatively the same temperature dependence of critical magnetic field and are in good agreement with the theoretical values at all the temperatures. The critical magnetic field vanishes at $T_c\simeq 27$~K that corresponds to the formation of gapless states, in which the system mimics massless Dirac fermions~\cite{Q3}. We formally extend our comparison toward negative values of $B_c$, which correspond to the formation of the NI state. Note that the error in the determination of $B_c$ from $\sigma_{xy}= \pm e^2/2h$ increases with temperature due to broadening of the trace of the zero-mode LL, originating from the \emph{H}1 subband as it is discussed above. Thus, our results shown in Fig.~\ref{Fig:4} are eloquent proof of the temperature-induced topological transition in HgTe QWs.

We note that the size of Hall bar in our sample largely exceeds the typical spin relaxation length ($\sim1$ $\mu$m)~\cite{Q15}. Additional four probes resistance measurements (see section~E in~\cite{SM}) evidence that although the edge states exist at $B<B_c$~\cite{Q27,Q28}, they do not contribute significantly to the conductivity of our sample.

Finally, we discuss the role of spin-orbital corrections resulting from bulk inversion asymmetry (BIA)~\cite{Q21} of zinc-blend crystals and interface inversion asymmetry (IIA)~\cite{Q22} on the behaviour of zero-mode LLs. Both reasons induce the anticrossing of zero-mode LLs in the vicinity of $B_c$. So far, fingerprint of such anticrossing was observed only by far-infrared LL spectroscopy~\cite{Q13,Q8,Q9,Q10} focused on LL transitions involving the zero-mode LLs.

We note that spin-orbital corrections resulting from structure inversion asymmetry (SIA) of the QW profile does not lead to anticrossing behaviour of zero-mode LLs. Moreover, since we do not observe any beatings of SdH oscillations induced by SIA at the gate voltages presented in Fig. 3, we conclude that SIA-induced corrections are small in our sample. The latter is also confirmed by good agreement between our experimental data and calculations performed for the symmetrical QW profile.

The linear dependence of energies of the zero-mode LLs and their experimental traces on magnetic field far from $B_c$ enables us to recalculate the anticrossing gap $\Delta$ measured in magnetooptics~\cite{Q8,Q9,Q10} into the values of gate voltage, expected in our sample. For instance, by using theoretical band-gap of 7~meV at 1.7~K and $B=3$~T (see Fig.~\ref{Fig:2}) and its experimental value $\Delta V_G\simeq0.7$ V represented in the scale of $V_g$ (see Fig 3(e)), we find that the experimental values of $\Delta\simeq$4-5 meV transform into 0.4--0.5~V, which should be observed for our sample. As it is seen from Fig.~\ref{Fig:3}, the width of experimental traces of zero-mode LLs in the vicinity of $B_c$, which may be interpreted as a manifestation of the LL anticrossing, is significantly lower than the mentioned values. The latter is also consistent with previous magnetotransport results~\cite{Q2,Q3}.


As discussed in~\cite{Q19}, a possible explanation of the large anticrossing gap in the vicinity of $B_c$ measured by far-infrared LL spectroscopy can be based on electron-electron interaction effect. The latter may largely influence close energies of LL transitions~\cite{Q23,Q24,Q25} as we deal with strongly non-parabolic 2D system for which Kohn's theorem~\cite{Q26} does not hold.

On the other hand, both BIA and IIA induces a spin splitting of both electron-like and hole-like states at non-zero $\mathbf{k}$ in the symmetrical QWs. If the spin splitting is strong enough, it results in the beatings arising in SdH oscillations. However, these beatings have never been observed in symmetrical HgTe QWs at low electron concentration. This is also consistent with the small strength of BIA and IIA terms, evaluated for our sample.

Finally, the presence of BIA and IIA terms induces the optical transitions between two branches of helical edge states~\cite{Q30}. If both terms are small only spin-dependent transitions between edge and bulk states are allowed~\cite{Q31}. Very recent accurate measurements of a circular photogalvanic current in HgTe QWs~\cite{Q29} have revealed the optical transitions between the edge and bulk states only. The latter also indicates the small effects of BIA and IIA terms in HgTe QWs.

In conclusion, we have directly observed a temperature-induced topological transition between NI and TI phases. By plotting experimental LL fan charts at different temperatures, we have accurately extracted values of critical magnetic field at different temperatures. Following experimental phase diagram, we have determined a critical temperature $T_c\simeq27$~K, at which $B_c$  vanishes. Our experimental results are in good agreement with realistic band structure and LLs calculations based on the eight-band Kane Hamiltonian with temperature dependent parameters.


\begin{acknowledgments}
~\\~The authors gratefully thanks Z. D. Kvon from the Institute of Semiconductor Physics (Siberian Branch, Russian Academy of Sciences) for processing of the samples. Part of this work was supported by the Languedoc-Roussillon region via the "Gepeto Terahertz platform", by Era.Net-Rus Plus project "Terasens", by the CNRS through "Emergence project 2016" and LIA "TeraMIR", by MIPS department of Montpellier University through the "Occitanie Terahertz Platform" and the ARPE project "Terasens", by and by the French Agence Nationale pour la Recherche (Grant No. ANR-16-CE09-0016 and Dirac3D project). Theoretical calculations were performed in the framework of project 16-12-10317 provided by the Russian Science Foundation. The growth of the samples were supported by the Russian Foundation for Basic Research (grants 15-52-16017, 18-52-16008). S.~S.~Krishtopenko and A.~M.~Kadykov acknowledge the Russian Ministry of Education and Science (MK-1136.2017.2 and SP-5051.2018.5).
\end{acknowledgments}


%

\newpage
\clearpage
\setcounter{equation}{0}
\setcounter{figure}{0}
\setcounter{table}{0}
\renewcommand{\thefigure}{S\arabic{figure}}

\onecolumngrid
\section*{Supplemental Materials}
\maketitle
\onecolumngrid

\subsection{A. Hall carrier density {\it vs} gate voltage}

\begin{figure}
 \includegraphics[width=0.7\linewidth]{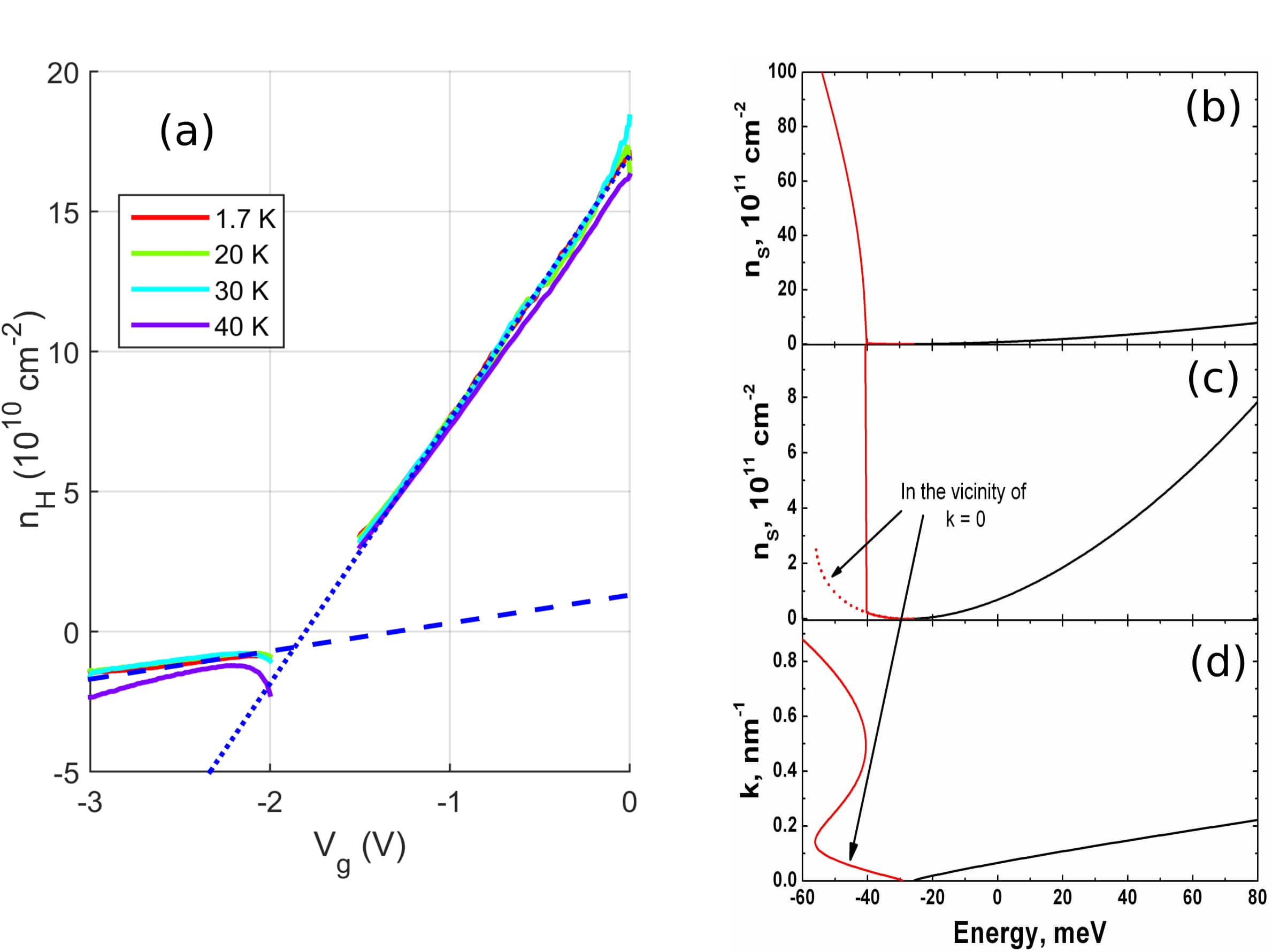}
	\caption{(a) Hall concentration $n_H$ as a function of the gate voltage $V_g$, at different temperatures $T=$ 1.7 K, 20 K, 30 K and 40 K. The effective capacitance is given by the thin dotted blue line (electron side) and thin dashed blue line (hole side).(b) Density of states (DoS) for the conduction band (black line) and valence band (red line). (c) the two DoS contributions corresponding to holes close to $k = 0$ (dashed red line) and $k \simeq 0.5$ nm$^{-1}$ (solid red line) are separated. (d) band structure evidencing the side maxima in the valence band.}
	\label{fig:nH}
\end{figure}

The Hall carrier density $n_H$ was first determined at various temperatures as a function of the gate voltage $V_g$. No significant difference in the carrier density was found along the Hall bar.
Fig.~\ref{fig:nH}a shows $n_H$ estimated from the Hall resistance $R_{28,13}$.
The geometric capacitance is theoretically equal to $C_g=$ $1.1 \times 10^{11}$ cm$^{-2} \cdot$eV.
It is in good agreement with the observed slope
$\partial n_H/\partial V_g \simeq 9 \times 10^{10}$ cm$^{-2}$/V on the electron side.

By contrast,  the experimental  slope
is significantly lower than expected on the hole side,
for $V_g < -2$ V.
The hole density remains pinned close to $p \simeq$ 1--2 $\times 10^{10}$ cm$^{-2}$.
This experimental observation is in good agreement with
the band structure calculation.
Figure.~\ref{fig:nH}(b-d) shows that
on the hole side, 10 meV below the topological band gap,
a valence band side maximum is present which can accumulate more than $10^{13}$ cm$^{-2}$ holes.
By contrast, holes with a small momentum $k$ are also present, but with a much smaller density.
These holes have a higher velocity and are expected to be much more mobile.
The holes trapped in the side maximum have a too small mobility to be detected in the Hall coefficient,
which is given only by the contribution of the holes close to $k=0$. However,
the overall large hole density
reduces significantly the gate efficiency and pins the Fermi  level approximately 10 meV below the band gap.

\subsection{B. Interpreting data at higher magnetic field}

\begin{figure}
\includegraphics[width=0.8\linewidth]{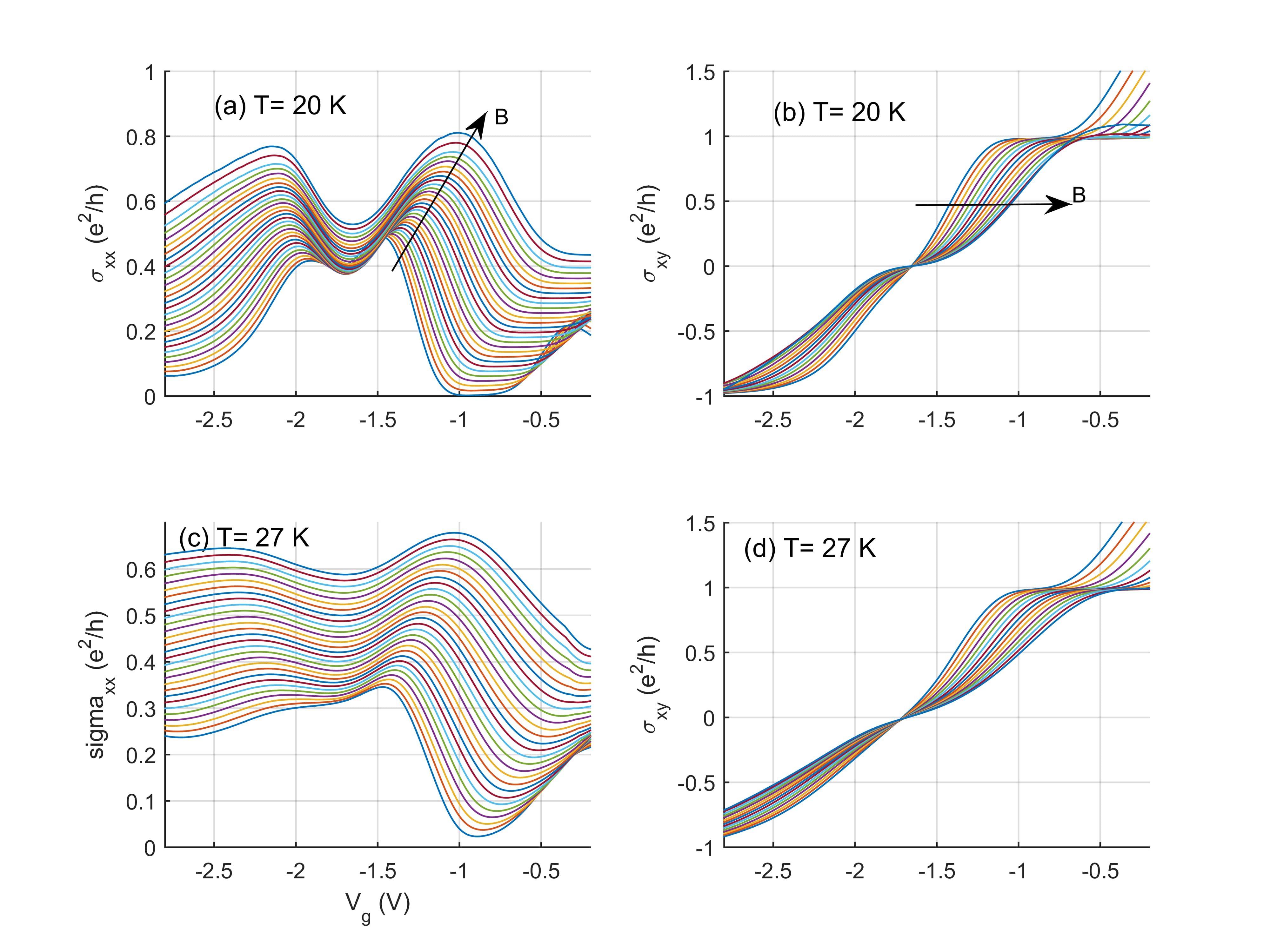}
\caption{Longitudinal and transverse conductivities at large $B$, $B=$ 3.2--6~T and at two temperatures $T=20$~K and $T= 27$~K. A vertical offset has been added to $\sigma_{xx}$ for clarity.}
	\label{fig:data1}
\end{figure}

Figure~\ref{fig:data1} shows the evolution of the conductivities at large magnetic fields, $B=$ 3.2 -- 6 T for two temperatures. Two maxima are clearly distinguishable in the longitudinal conductitivities. When the temperature increases, the voltage difference between the two maxima also increases. This is the key signature of the progressive transformation of the TI gap into a NI gap. Simultaneously, the
gate voltage between the two conductivities corresponding to $\sigma_{xy}= \pm e^2/2h$ also increases with $B$ and $T$, see panels (b) and (d).

Figure 3(e,f) (main text) shows colormaps of the longitudinal
magnetoresistance measured at $T=1.7$ K and $T=20$ K.
At $B=0$ T, a resistance maximum is identified around $V_g \simeq -1.8$ V, which corresponds to the topological gap.
In increasing magnetic field, Landau levels are clearly distinguishable on the electron side ($V_g > -1.8 $ V).
On the hole side  ($V_g < -1.8$ V),
LLs are also observed, with a very small dependence on $V_g$.
The carrier density extracted from the $1/B$ periodicity of these LLs is in agreement with the Hall density obtained at low field.
Following the above arguments, the LLs observed for $V_g < -1.8$ V are associated to holes close to $k=0$, whereas the LLs associated to the hole side maximum of the valence band are not observed.



Remarkably, the resistance shows a clear saddle point around $B \simeq  1$ T and $V_g \simeq  -1.8$ V.
This saddle point originates from the crossing of the two zero-modes LLs and its observation
is possible because of the presence of disorder.
To show this, let us assume that the LLs have a Lorentzian broadening $\Gamma$. 
Because  the energy difference between adjacent LLs is comparable to $\Gamma$,
the LLs overlap.
At the maximum of the density of states of the LL of index $N_0$, the filling factor $\nu$  does not correspond to an half-integer. It is given by:
 \begin{equation}
 \nu = \sum_{N} \int_{\mp\infty}^{E_{N_0}}
 \frac{1} {\pi}
 \frac{\frac{1}{2}\Gamma}{(E-E_N)^2+ (\frac{1}{2}\Gamma)^2}dE
 \end{equation}
 which can be simplified as:

\begin{equation}
\nu = \sum_N  \left[\pm \frac{1}{2}+ \frac{1}{\pi}\arctan \left(2  \frac{E_{N_0}-E_N}{\Gamma}\right) \right]
\end{equation}
where $N$ is the LL index and $E_N$ is the energy of the $N^{th}$ LL as calculated theoretically.
The $\pm$ sign refers to LLs associated either to the conduction or the valence bands respectively.
The only fitting parameter is the LL broadening $\Gamma$.

\begin{figure}
\includegraphics[width=1.0 \linewidth]{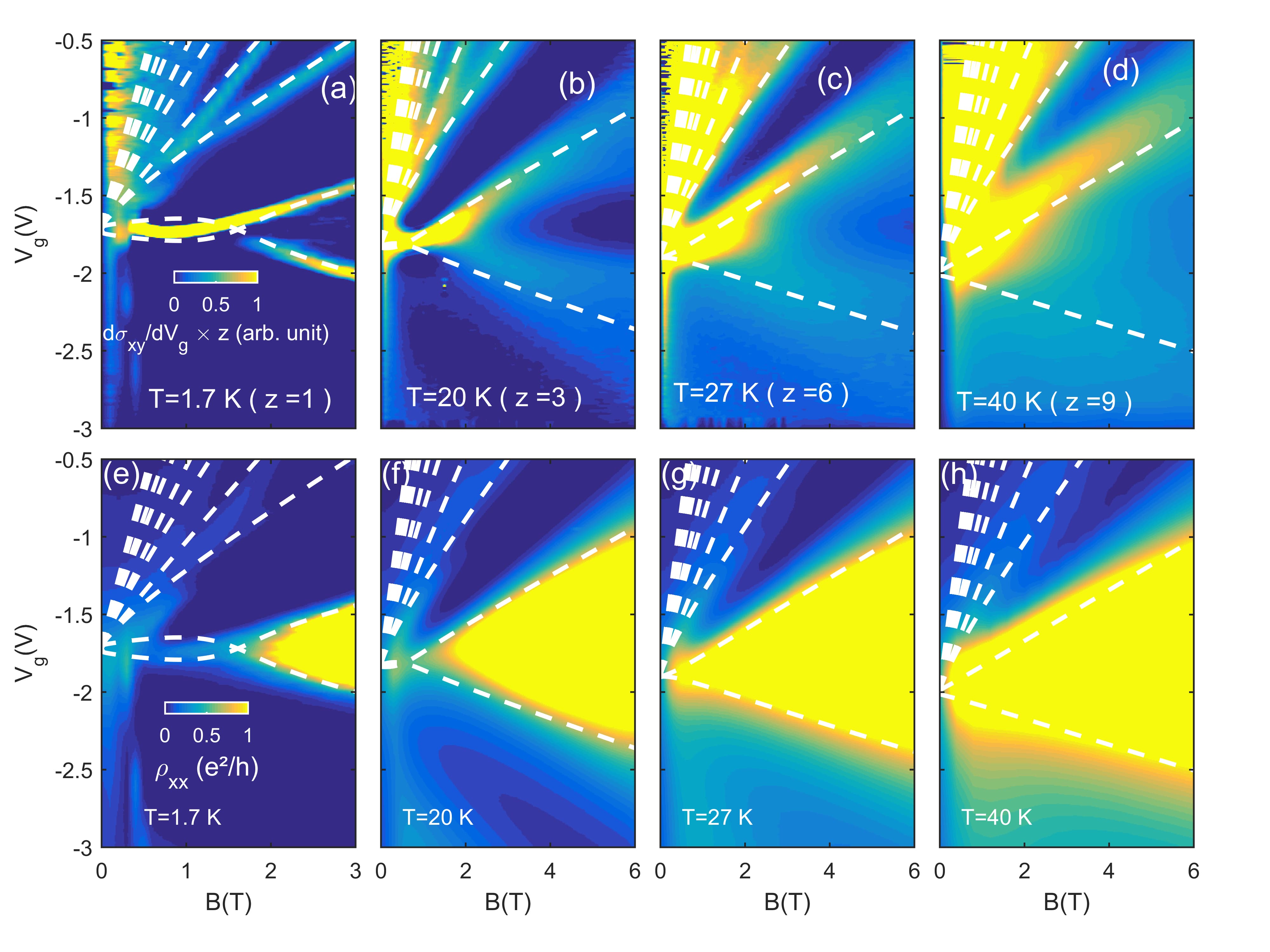}
	\caption{(a-d) Color maps of $\partial\sigma_{xy}/\partial V_g$, at 4 different temperatures. The superimposed white lines are the calculated LL positions in the $(V_g, B)$ plane. (e-h) Colormaps for the resistivity $\rho_{xx}$, at the same four temperatures. The white lines indicate the same calculated LL positions.}
	\label{fig:simul}
\end{figure}

The filling factor $\nu$ is directly related to the gate voltage by the geometric capacitance:
$C_\mathbf{geo} (V_g- V_0) = n e = \nu e^2 B/h$, where $V_0$ is a constant.
The quantum capacitance is negligible in our geometry.
From the above equations, the positions of the LLs can be calculated as a function of $V_g$ and $B$. The calculated position of the first LLs of the conduction band are shown
in Fig.~\ref{fig:simul}(e-h) in the $(B, V_g$) plane as dashed white lines,
superposed to $\rho_{xx}$, and
in Fig.~\ref{fig:simul}(a-d)  superposed to
the derivative of the experimental  transverse conductivity
$\partial\sigma_{xy}/\partial V_g$.
To obtain a good agreement
with data, we have chosen
$\Gamma= 4$ meV.

Most of the LLs follow straight lines, as expected if the LL overlap is negligible:
the LL position is then given by a constant $\nu$ close to a half-integer.
However, the evolution of the two zero mode LLs is much more complex, because these LLs  cross at the critical field $B_c$ where $\nu=0$.
Let us stress that the two zero mode LLs emerge from  the $E1$ and $H1$ bands, respectively,
which are separated by an energy gap of several meV. Nevertheless, from Fig.~\ref{fig:simul}, these two LLs seem to emerge from the same gate voltage at $B=0$ T.
The reason for this is that the gate voltage is proportional to the net carrier density. As our model does not include any disorder in the topological gap at $B=0$ T, the two LLs merge at $B=0$~T in the $(V_g, B)$ plane.

\subsection{C. Estimation of $B_c$}

\begin{figure}
\includegraphics[width=0.95\linewidth]{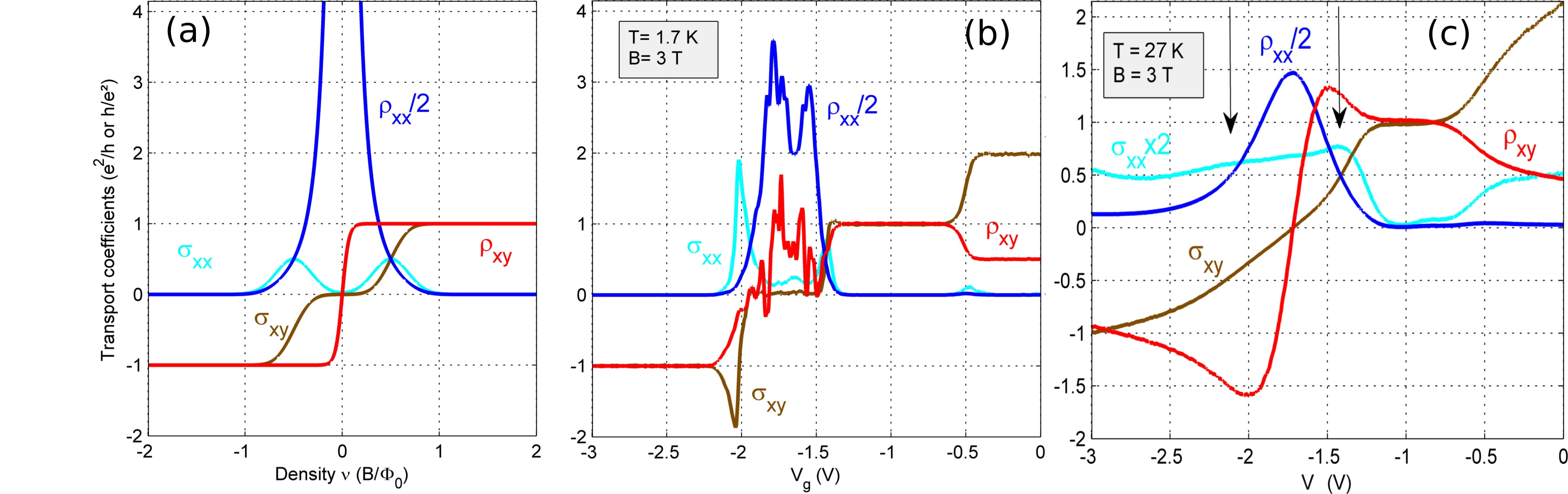}
	\caption{(a) Theoretical transport coefficients as a function of filling factor $\nu$.  (b) Experimental transport coefficients at $T = 1.7$ K, $B= 3$ T and (c) $T = 27$ K, $B= 3$ T, {\it versus} $V_g$. }
	\label{fig:S4}
\end{figure}
To get an estimate of $B_c$, the position of the two zero mode LLs must be determined accurately from the data.
Fig.~\ref{fig:S4}a sketches the theoretical evolution of both conductivity and resistivity when the filling factor is varied across the normal gap, at high $B$.
The LLs at $\nu= \pm 1/2$ correspond to
i) a maximum of the longitudinal conductivity $\sigma_{xx}$;
ii) a value $\pm e^2/2h$ of the transverse conductivity;
iii) a maximum of the derivative $\partial \sigma_{xy}/\partial V_g$;
iv) a value $h/e^2$ for the longitudinal resistivity $\rho_{xx}$, as far as the semicircle relation remains valid~\cite{Dykhne1994PRB,Shahar1997}.

Fig.~\ref{fig:S4}b and~\ref{fig:S4}c show the experimental transport coefficients
at $B= 3$ T, $T= 1.7$ K and $T= 27$~K, respectively.
The relations (i)-(iv) are indeed satisfied. however, at the highest temperature,
the maxima of $\sigma_{xx}$ (indicated by arrows) and $\partial \sigma_{xy}/\partial V_g$ almost disappeared, whereas $\rho_{xx}$ and $\sigma_{xy}$ are still quite close to their expected values.
%
%
Therefore the conditions ii) and iv) have been used preferentially to find the position of the LLs at magnetic fields above 2.2 T for different temperatures.
We have chosen $B=2.2$ T as a minimal threshold because this is the lowest magnetic field for which we still observe
$\rho_{xx}=h/e^2$ when  $\sigma_{xy}=\pm e^2/2h$ (with an uncertainty of 3\% ).

A linear extrapolation of these isolines $\rho_{xx}= h/e^2$ and
$\sigma_{xy}= \pm e^2/h$  toward low magnetic field gives two estimates of $B_c$. For each estimate, a mean value and a standard deviation ($1\sigma$) can be calculated by varying the magnetic field interval on which the linear fit is done.

\subsection{D. Conductivity mechanisms in the NI gap}

\begin{figure}
\includegraphics[width=0.7\linewidth]{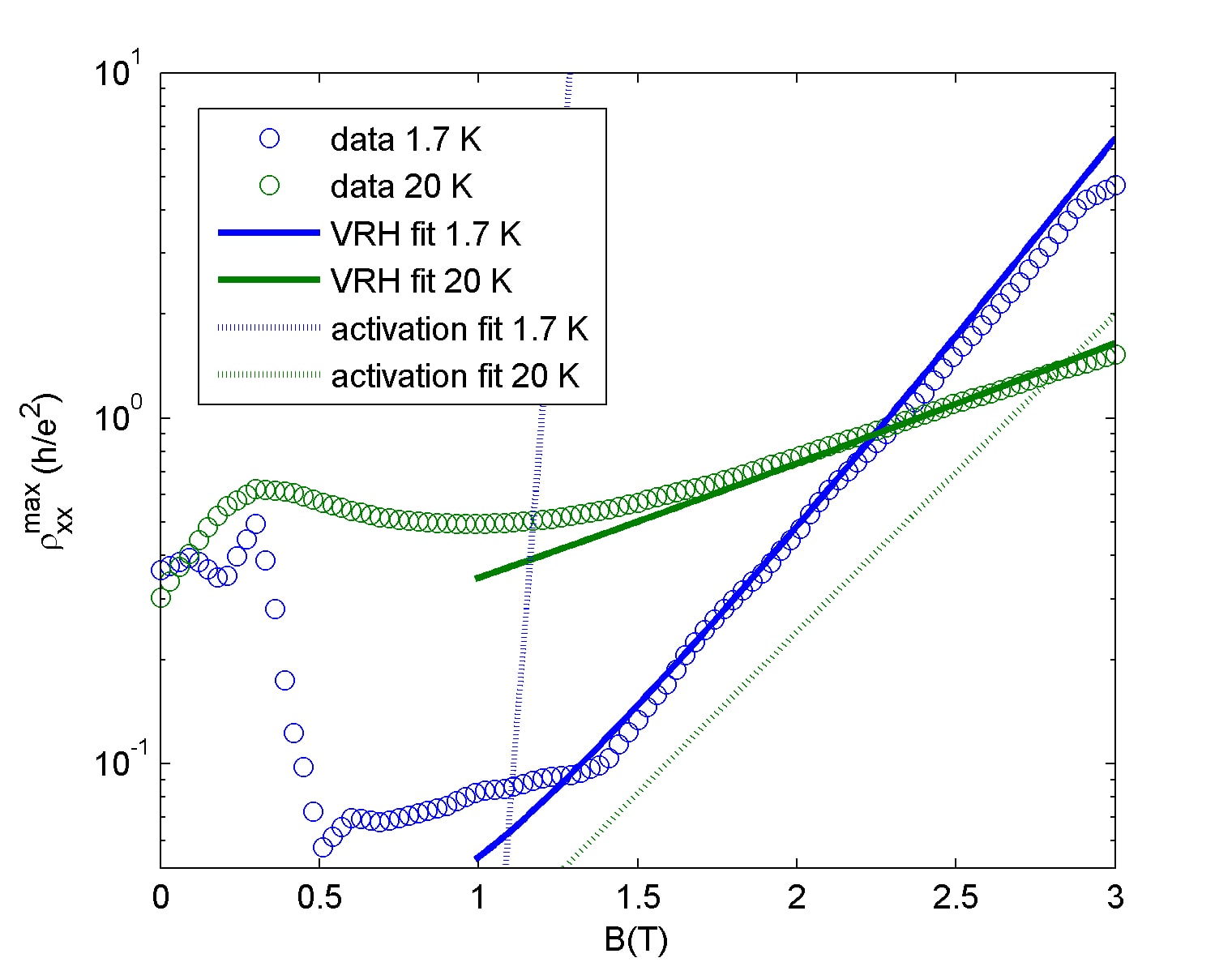}
	\caption{ Maximum of the magnetoresistance $\rho_{xx}^{max}(B)$ around $\nu \simeq 0$, at $T=1.7$ K (blue open circles) and $T=20$ K (green open circles). The blue and green solid lines are VRH fits of the magnetoresistance. The dashed lines correspond to an activation law $\rho_{xx} \propto \exp(-\Delta E/2k_BT) $ which cannot fit the data.}
	\label{fig:normalgap}
\end{figure}

Figure~\ref{fig:normalgap} shows the maximum $\rho_{xx}^\mathrm{max}$ of the magnetoresistance around $\nu = 0$
 ($\rho_{xx}^\mathrm{max} = \max\limits_{-2 V < V_g < -1.5 V } \rho_{xx}$) from $B=0$~T to $B=3$~T, at two temperatures.
At $T=1.7$~K, the resistance dip is rather large (from $B=0.5$~T to 1.5~T) with a minimum around $B=0.5$~T. The resistance falls below the expected theoretical value of $h/e^2$,
possibly because of additional parallel conduction.
At higher temperature $T=20$ K, the resistance dip is better defined and appears at $B=1$~T.
As the $\rho_{xx}$ minimum goes to higher $B$ when $T$ increases,
its position cannot be directly related to the crossing of the two zero-mode LLs.
By contrast, the resistivity at $B>1.5$~T can be easily explained,
qualitatively and quantitatively.
Above $B \simeq  1.5 $ T,
the magnetoresistance increases almost exponentially.
As the gap is theoretically proportional to $B$, it means that the resistance increases exponentially with the gap.

First, a simple activation law  cannot fit satisfactorily the data because the experimental temperature dependence
of the resistivity is too small.
The dotted lines in Fig.~\ref{fig:normalgap} show a tentative fit  $ \rho_{xx} = \rho_0 \exp (-\Delta E/k_BT)$,
where $\Delta E(B)$ is half the theoretical gap.
Obviously, the fit is very poor.

Second, For two-dimensional systems in the quantum Hall regime,
variable range hopping (VRH) is often the dominant conduction mechanism.
For VRH with a Coulomb gap, in the so-called Efros-Shklovskii VRH, (EF-VRH),
the conductivity varies as $\sigma \propto (\sigma_0/ T) \times \exp(-\sqrt{T_0/T})$.
Here, $T_0$ is a fit parameter, which depends on the localization length $\xi$ as:
$
k_B T_0 = C e^2 / 4 \pi \epsilon_r \epsilon_0 \xi,
$
where
$\epsilon_0$ is the vacuum dielectric constant,
$\epsilon_r \simeq 10.3$ is the relative dielectric constant,
and $C \simeq $ 6.2 is a constant.
From the scaling theory of the QHE, $\xi$ is expected to vary as
$\xi \propto \Delta E^{-\gamma}$, where $\gamma \simeq 2.3$,
$\Delta E$ is the energy separation between the electrochemical potential
and the LL delocalized states.
Assuming $\Delta E$ is half the theoretical gap between the two zero mode LLs,
EF-VRH can fit satisfactorily
both $B$ and $T$-dependence of the magnetoresistance,
see Fig.~\ref{fig:normalgap}.
From the ES-VRH fit, one can tentatively extract the localization length:
$\xi \simeq 400$ nm at $B = 1.5$ T and $\xi \simeq 100$ nm at $B=3$ T.
%

Our results and our analysis suggest that VRH alone
describes the transport not only on the Hall plateaus but also
at the plateau transition, as originally suggested by Polyakov-Shklovskii~\cite{PolyakovPRL1993}
and as recently observed in graphene~\cite{Bennaceur2012PRB}. The good understanding of the transport properties around the normal gap allows us to use these data in confidence to get a better estimation of $B_c$.

\subsection{E. Conductivity mechanisms in the TI gap}

\begin{figure}
\includegraphics[width=0.7\linewidth]{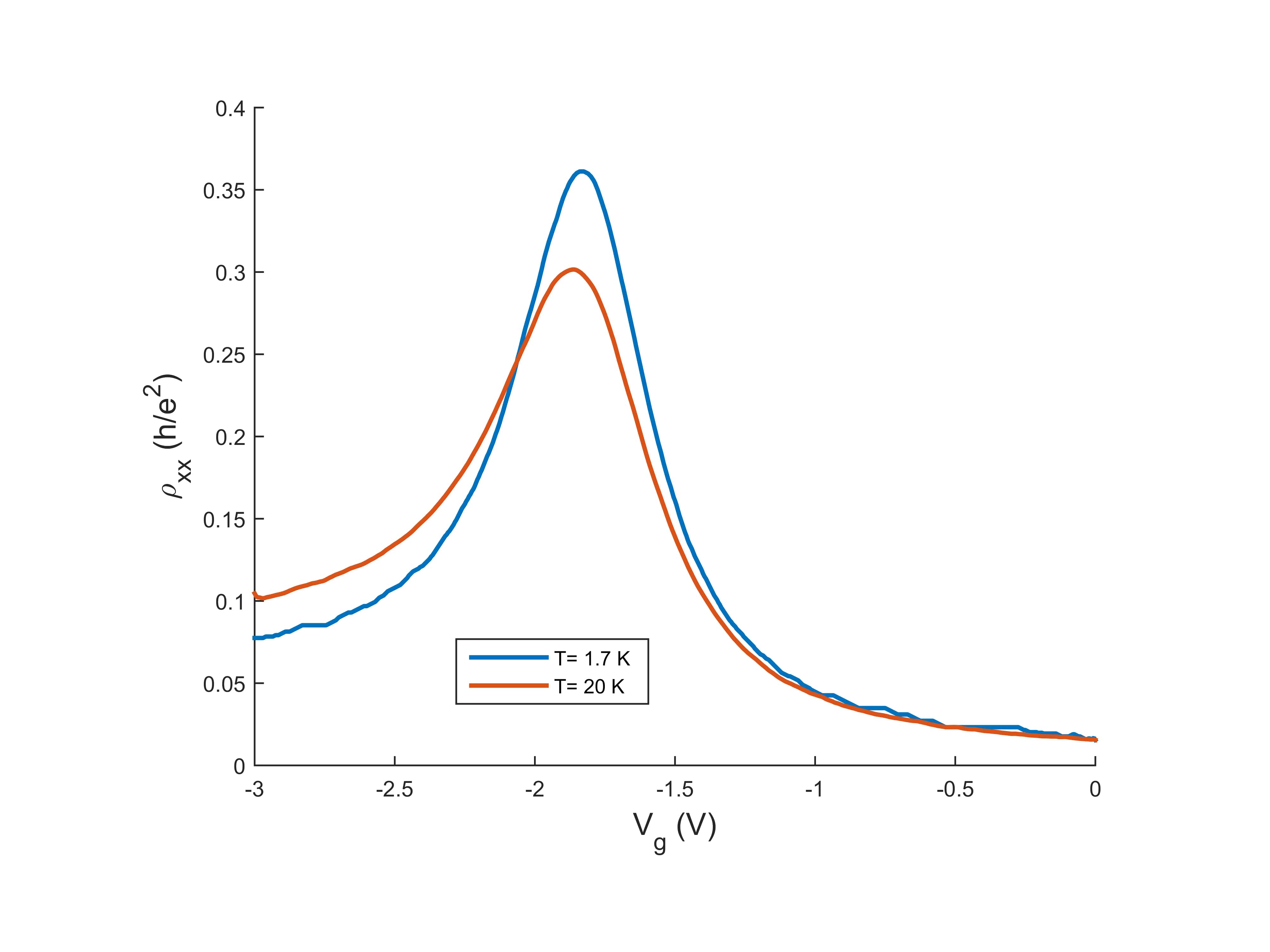}
	\caption{ Resistivity as a function of gate voltage, at two different temperatures and at $B=0 $ T.}
	\label{fig:data2}
\end{figure}

Conductivity mechanisms in the TI gap are more difficult to determine. Fig.~\ref{fig:data2} shows $\rho_{xx}(V_g)$ at $B=0$~T, $T= 1.7$~K and $T= 20$~K. A small insulating behavior is visible for the resistance maximum, as expected for bulk conductivity. As most reports point towards a spin relaxation length of $\sim$1~$\mu$m~\cite{sm3}, it would be very unlikely to observe the conductivity of the topological edge states in our large sample. Let us study here the influence of counter propagating edge states through non local transport measurements. Let us first assume that the bulk conductivity is completely negligible, and that backscattering is forbidden between the edge states because of spin conservation. At each metallic contact, the two counter-propagating states equilibrate. The four-probe resistance is then given by:
\begin{equation}
R^\infty=(n+1)\times \frac{N_u+1}{N_u+N_l+2} \times h/e^2
\label{eq:nl1}
\end{equation}
where $N_u$, $N_l$ and $n$ are the respective numbers of floating contacts along the upper edge, along the lower edge and between the two (local or non-local) probes located on the lower edge. The $\infty$ exponent indicates that no backscattering takes place along the two edge states in the QW.

If backscattering takes places along the edge states with a typical length
$\lambda$,  the upper equation becomes:
\begin{equation}
R^\lambda \simeq \frac{L}{\lambda} \times \frac{ L_u}{L_u+L_l }×h/e^2
\label{eq:nl2}
\end{equation}
where $L_u$, $L_l$ are the lengths of the upper and lower edges and $L$
is the distance between the two voltage probes.

When the current is injected between contacts 12 and 4 and voltage measured between contacts 1 and 3 (see Fig.~1 in the main text), $N_u$=5, $N_l$=3, $n$=1. This gives
$R_{1-3,12-4}^\infty=1.2×h/e^2$ and $R_{1-3,12-4}^{500\mu m} \approx 0.7 h/e^2$ for $L =\lambda = 500~\mu$m. The last estimation for $\lambda$ is very large.
More realistic values would considerably increase the resistance.
Therefore, the two above equations give the minimal value for the non local resistance  $R_{1-3,12-4}$ when the conduction is driven only by edge states. Experimentally, we found $R_{1-3,12-4}= 0.001~h/e^2$ at $T=1.7$ K, $B=0$ T and $V_g= -1.8$ V (the resistance maximum). This is three orders of magnitude smaller than the previous expectations based on edge states.

By contrast, the classical van der Pauw current spreading also gives non local resistances. For the non- local configuration, this ohmic contribution is given by the formula:
\begin{equation}
R_{1-3,12-4}^{ohmic}= \frac{4}{\pi} \rho_{xx}  \exp⁡(-\pi L/W)
\label{eq:nl3}
\end{equation}
As  $L/W=2$ for this pair of contacts, using in this formula the measured value $R_{1-3,12-4}= 0.001 h/e^2$ , we get $\rho_{xx} \simeq   0.4~ h/e^2$,
which is very close to the value given by the local experimental resistance  $R_{28,34}=L/W \times 0.36~h/e^2$. We conclude that the resistance is largely dominated by the bulk contribution, in both local and non-local configurations (for completeness, Eqs.~\ref{eq:nl1} and~\ref{eq:nl2} give  $R_{28,34}^\infty=0.5~h/e^2$ and
$R_{28,34}^{1\mu m}=1400~h/e^2)$.

Finally, a linear magnetoresistance appears at $V_g \simeq -1.8$ V, around $T= $20--30 K and below $B = 0.5$ T.
This gives rise to an apparent dip in the resistance in Fig. 3(g-h) of the main text.
We speculate that this behavior is due to
a quantum magnetoresistance effect, as
proposed for materials with zero bandgap and linear dispersion~\cite{Abrikosov1998}.
In this scenario, the linear magnetoresistance is another signature of the
closure of the gap around 20--30 T.

\subsection{F. Bernevig-Hughes-Zhang Hamiltonian}
To describe the band inversion in HgTe/Cd(Hg)Te QWs, in which the first electron-like \emph{E}1 and hole-like \emph{H}1 subbands are very close in energy, one can also use the effective Dirac-type Bernevig-Hughes-Zhang (BHZ) Hamiltonian~\cite{sm1}, proposed for the electronic states in the vicinity of the $\Gamma$ point of the Brillouin zone. The latter is derived from the Kane Hamiltonian~\cite{sm2}, which includes $\Gamma_6$, $\Gamma_8$, $\Gamma_7$ bulk bands with the confinement effect. Within the representation defined by the basis states $|$\emph{E}1,+$\rangle$, $|$\emph{H}1,+$\rangle$, $|$\emph{E}1,-$\rangle$, $|$\emph{H}1,-$\rangle$, the effective 2D Hamiltonian has the form:
\begin{equation}
\label{eq:BHZ1}
H_{2D}(\mathbf{k})=\begin{pmatrix}
H_{\mathrm{BHZ}}(\mathbf{k}) & 0 \\ 0 & H_{\mathrm{BHZ}}^{*}(-\mathbf{k})\end{pmatrix},
\end{equation}
where asterisk stands for complex conjugation, $\mathbf{k}=(k_x,k_y)$ is the momentum in the QW plane, and
\begin{equation}\label{eq:BHZ2}
\hat{H}_{\mathrm{BHZ}}(\textbf{k})=\epsilon(\textbf{k})\mathbf{I}_2+\sum_{i=1}^3 d_i(\textbf{k})\sigma_i,
\end{equation},
\begin{equation*}
d_1+id=A(k_x+ik_y)=Ak_{+},~~~~d_3=M-B(k_x^2+k_y^2),~~~~\epsilon=C-D(k_x^2+k_y^2).
\end{equation*}
Here, $\mathbf{I}_2$ is a 2$\times$2 unit matrix, $\sigma_a$ are the Pauli matrices, $\epsilon_{\mathbf{k}}=C-D(k_x^2+k_y^2)$, $d_1(\mathbf{k})=-Ak_x$, $d_2(\mathbf{k})=-Ak_y$, and  $d_3(\mathbf{k})=M-B(k_x^2+k_y^2)$. The structure parameters $C$, $M$, $A$, $B$, $D$ depend on $d$, Cd concentration in the barrier and external conditions (pressure, temperature, etc.). The sign of mass parameter $M$ describes inversion between \emph{E}1 and \emph{H}1 subbands: $M>0$ corresponds to a trivial state, while, for a QSHI state, $M<0$. We note that $H_{2D}(\mathbf{k})$ has a block-diagonal form because the terms, which break inversion symmetry and axial symmetry around the growth direction, are being neglected~\cite{sm2}. The latter is a quite good approximation for symmetric HgTe QWs.

To calculate Landau levels (LLs) in the presence of an external magnetic field $\mathbb{B}$ oriented perpendicular to the QW plane, one should make the Peierls substitution, $\hbar\mathbf{k}\longrightarrow\hbar\mathbf{k}-\frac{e}{c}\mathbf{A}$, where in the Landau gauge $\mathbf{A}=\mathbb{B}(y,0)$. Additionally, we add the Zeeman term in the Hamiltonian
\begin{equation}
\label{eq:BHZ3}
H_{Z}=\dfrac{1}{2}\mu_B\mathbb{B}\begin{pmatrix}
g_e & 0 & 0 & 0 \\
0 & g_h & 0 & 0 \\
0 & 0 & -g_e & 0 \\
0 & 0 & 0 & -g_h\end{pmatrix},
\end{equation}
where $\mu_B$ is the Bohr magneton, $g_e$ and $g_h$ are the effective (out-of-plane) g-factors of the \emph{E}1 and \emph{H}1 subbands, respectively.

Solving the eigenvalue problem for the upper block of $H_{2D}(\mathbf{k})+H_{Z}$, the LL energies $E^{(+)}_n$ are found analytically~\cite{sm3}:
\begin{equation*}
E^{(+)}_n=C-\dfrac{2Dn+B}{a_B^2}+\dfrac{g_e+g_h}{4}\mu_B\mathbb{B}
\pm\sqrt{\dfrac{2nA^2}{a_B^2}+\left(M-\dfrac{2Bn+D}{a_B^2}+\dfrac{g_e-g_h}{4}\mu_B\mathbb{B}\right)^2}, \text{~~~~~~~for $n\geq1$}
\end{equation*}
\begin{equation}
\label{eq:BHZ4}
E^{(+)}_0=C+M-\dfrac{D+B}{a_B^2}+\dfrac{g_e}{2}\mu_B\mathbb{B} \text{~~~~~~~for $n=0$}.
\end{equation}
For the lower block upper block of $H_{2D}(\mathbf{k})+H_{Z}$, the LL energies $E^{(-)}_n$ are written as
\begin{equation*}
E^{(-)}_n=C-\dfrac{2Dn-B}{a_B^2}-\dfrac{g_e+g_h}{4}\mu_B\mathbb{B}
\pm\sqrt{\dfrac{2nA^2}{a_B^2}+\left(M-\dfrac{2Bn-D}{a_B^2}-\dfrac{g_e-g_h}{4}\mu_B\mathbb{B}\right)^2}, \text{~~~~~~~for $n\geq1$}
\end{equation*}
\begin{equation}
\label{eq:BHZ5}
E^{(-)}_0=C-M-\dfrac{D-B}{a_B^2}-\dfrac{g_h}{2}\mu_B\mathbb{B} \text{~~~~~~~for $n=0$}.
\end{equation}
Here $a_B$ is the magnetic length given by $a_B^2=\hbar c/e\mathbb{B}$. The LLs with energies $E^{(+)}_0$ and $E^{(+)}_0$ are called the \emph{zero-mode} LLs~\cite{sm3}. They split from the edge of \emph{E}1 and \emph{H}1 subbands and tend toward conduction and valence band, respectively. The crossing of the zero-mode LLs occurs at a critical magnetic field $B_c$ (see Fig.~2 in the main text), above which the inverted band ordering is transformed into the trivial one~\cite{sm3}. One can show that $B_c>0$ for HgTe QWs with inverted band structure ($M<0$).

\begin{figure}
	\includegraphics[width=0.95\linewidth]{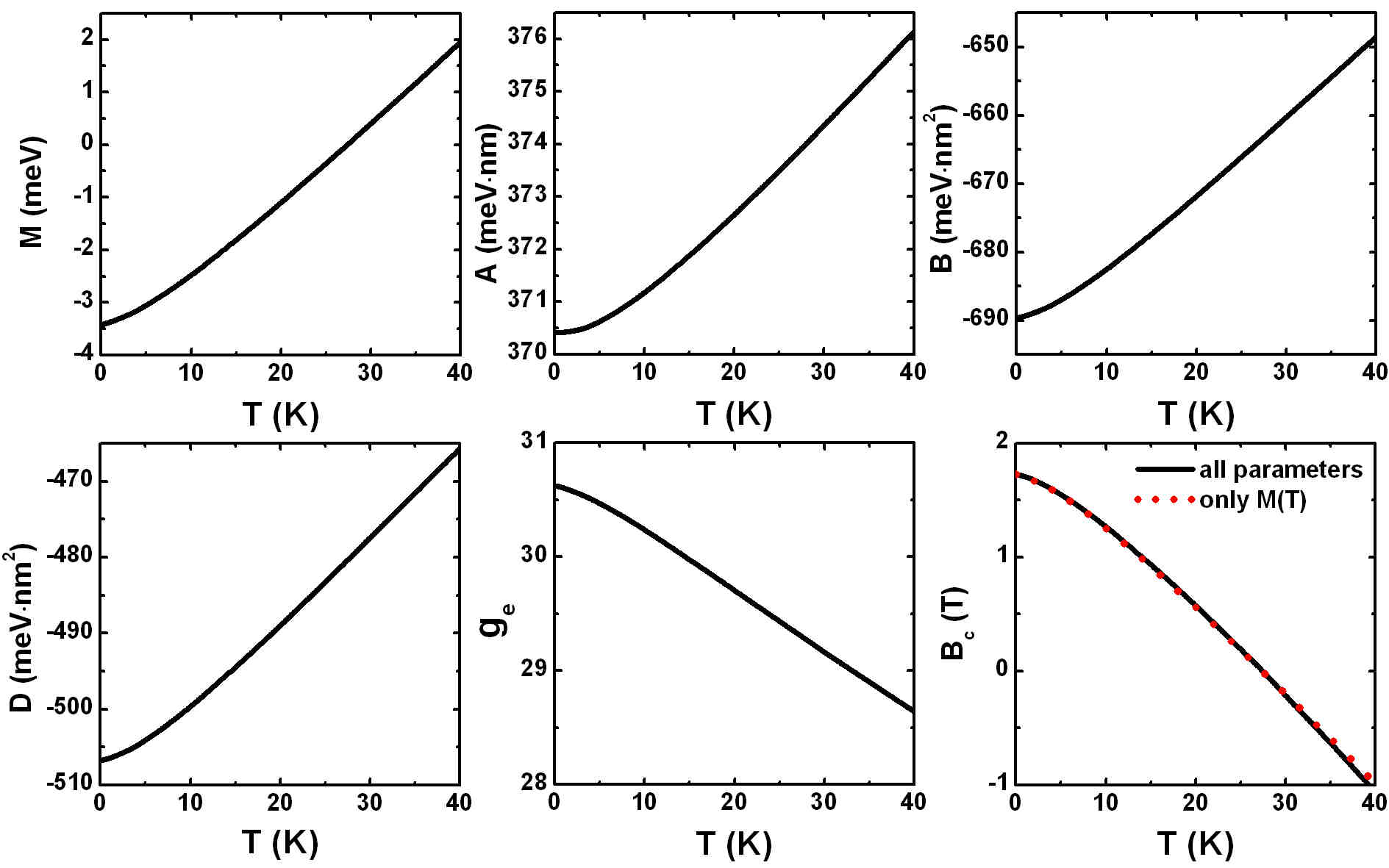} 
	\caption{Dependence of the parameters of the BHZ Hamiltonian and critical magnetic field $Bc$ on temperature, calculated for our sample (see the main text). We note that effective g-factor of the \emph{H}1 subband does not depend on temperature, $g_h=-1.21$. The latter is caused by independence of Luttinger parameters $\gamma_1$, $\gamma_2$, $\gamma_3$ and $\kappa$ on $T$ in the Kane Hamiltonian~\cite{sm2}.}
	\label{fig:S6}
\end{figure}

By using the 8-band Kane Hamiltonian, accounting interaction between the $\Gamma_6$, $\Gamma_8$ and $\Gamma_7$ bands in zinc-blend materials, with temperature-dependent parameters\cite{sm2} and by applying the procedure, described in~\cite{sm4}, we have calculated the values of $A$, $B$, $C$, $D$, $g_e$, $g_h$ and $M$ for our sample at different temperatures (see Fig.~\ref{fig:S6}).

Figure~\ref{fig:S6} also shows a critical magnetic field as a function of temperature $T$. The solid black curve represents the calculations performed within the Kane Hamiltonian~\cite{sm2}. We note that the calculation of $B_c$ based on Eqs.~(\ref{eq:BHZ4}) and~(\ref{eq:BHZ5}), performed by taking into account temperature dependence of all parameters in the BHZ Hamiltonian, coincides with the values within the Kane model. The dotted red curve corresponds to the calculation of $B_c$ within the BHZ model if we take into account only temperature dependence of the mass parameter $M$. The small difference between the curves means that temperature dependence of $B_c$ is mostly caused by temperature dependence of the band gap in our sample (see also Fig.~2 in the main text), while the temperature effect on the dispersion of the zero-mode LLs is negligible.

%

\end{document}